# On the Parity-Check Density and Achievable Rates of LDPC Codes

Gil Wiechman    Igal Sason

Technion – Israel Institute of Technology
Haifa 32000, Israel
{igillw@tx, sason@ee}.technion.ac.il

**Abstract**

The paper introduces new bounds on the asymptotic density of parity-check matrices and the achievable rates under ML decoding of binary linear block codes transmitted over memoryless binary-input output-symmetric channels. The lower bounds on the parity-check density are expressed in terms of the gap between the channel capacity and the rate of the codes for which reliable communication is achievable, and are valid for every sequence of binary linear block codes. The bounds address the question, previously considered by Sason and Urbanke, of how sparse can parity-check matrices of binary linear block codes be as a function of the gap to capacity. The new upper bounds on the achievable rates of binary linear block codes tighten previously reported bounds by Burshtein *et al.*, and therefore enable to obtain tighter upper bounds on the thresholds of sequences of binary linear block codes under ML decoding. The bounds are applied to low-density parity-check (LDPC) codes, and the improvement in their tightness is exemplified numerically.

## 1 Introduction

Error correcting codes which employ iterative decoding algorithms are now considered state of the art in the field of low-complexity coding techniques. These codes closely approach the capacity limit of many standard communication channels under iterative decoding. In [4], Khandekar and McEliece have suggested to study the encoding and decoding complexities of ensembles of iteratively decoded codes as a function of their gap to capacity. They conjectured that if the achievable rate under message-passing iterative (MPI) decoding is a fraction $1 - \varepsilon$ of the channel capacity, then for a wide class of channels, the encoding complexity scales like $\ln \frac{1}{\varepsilon}$ and the decoding complexity scales like $\frac{1}{\varepsilon} \ln \frac{1}{\varepsilon}$. The only exception is the binary erasure channel (BEC) where the decoding complexity behaves like $\ln \frac{1}{\varepsilon}$ (same as encoding complexity) because of the absolute reliability of the messages passed through the edges of the graph (hence, every edge can be used only once during the iterative decoding process).

Low-density parity-check (LDPC) codes are efficiently encoded and decoded due to the sparseness of their parity-check matrices. In his thesis [2], Gallager proved that right-regular LDPC codes (i.e., LDPC codes with a constant degree ($a_R$) of the parity-check nodes) cannot achieve the channel capacity on a BSC, even under optimal ML decoding.

This inherent gap to capacity is well approximated by an expression which decreases to zero exponentially fast in $a_R$. Richardson *et al.* [10] have extended this result, and proved that the same conclusion holds if $a_R$ designates the *maximal right degree*. Sason and Urbanke later observed in [12] that the result still holds when considering the *average right degree*. Gallager's bound [2, Theorem 3.3] provides an upper bound on the rate of right-regular LDPC codes which achieve reliable communications over the BSC. Burshtein *et al.* have generalized Gallager's bound for general ensembles of LDPC codes transmitted over memoryless binary-input output-symmetric (MBIOS) channels [1], and the work in [12] relies on their generalization.

Consider the number of ones in a parity-check matrix which represents a binary linear code, and normalize it per information bit (i.e., with respect to the dimension of the code). This quantity (which will be later defined as the *density* of the parity-check matrix) is equal to $\frac{1-R}{R}$ times the average right degree of the bipartite graph that represents the code, where $R$ is the rate of the code in bits per channel use. In [12], Sason and Urbanke considered how sparse can parity-check matrices of binary linear block codes be, as a function of their gap to capacity (where this gap depends in general on the channel and on the decoding algorithm). An information-theoretic lower bound on the asymptotic density of parity-check matrices was derived in [12, Theorem 2.1] where this bound applies to every MBIOS channel and *every* sequence of binary linear block codes achieving a fraction $1 - \varepsilon$ of the channel capacity with vanishing bit error probability. It holds for an arbitrary representation of full-rank parity-check matrices for these codes, and is of the form $\frac{K_1 + K_2 \ln \frac{1}{\varepsilon}}{1-\varepsilon}$ where $K_1$ and $K_2$ are constants which only depend on the channel. Though the logarithmic behavior of this lower bound is in essence correct (due to a logarithmic behavior of the upper bound on the asymptotic parity-check density in [12, Theorem 2.2]), the lower bound in [12, Theorem 2.1] is *not* tight (with the exception of the BEC, as demonstrated in [12, Theorem 2.3], and possibly also the binary symmetric channel (BSC)). The derivation of the bounds in this paper was motivated by the desire to improve the results in [1, Theorems 1 and 2] and [12, Theorem 2.1] which are based on a two-level quantization of the log-likelihood ratio (LLR). They provide upper bounds on the achievable rates of LDPC codes over MBIOS channels, and lower bounds on their asymptotic parity-check density.

In [5], Measson and Urbanke derived an upper bound on the thresholds under maximum-likelihood (ML) decoding of LDPC ensembles when the codes are transmitted over the BEC. Their general approach relies on EXIT charts, having a surprising and deep connection with the maximum a posteriori (MAP) threshold due to the area theorem for the BEC. Generalized extrinsic information transfer (GEXIT) charts were recently introduced by Measson, Montanari, Richardson and Urbanke [6]; they form an extension of EXIT charts to general MBIOS channels, and in particular, they fulfill the area theorem (see [11, Section 3.4.10]). This conservation law enables one to get upper bounds on the thresholds of turbo-like ensembles under bit-MAP decoding. The bound was shown to be tight for the BEC [5], and was conjectured to be tight for general MBIOS channels.

A new method for analyzing LDPC codes and low-density generator-matrix (LDGM) codes under bit MAP decoding is introduced by Montanari in [7]. The method is based on a rigorous approach to spin glasses, and allows a construction of lower bounds on the entropy of the transmitted message conditioned on the received one. The calculation of this bound is rather complicated, and its complexity grows exponentially with the maximal right and left degrees (see [7, Eqs. (6.2) and (6.3)]); this imposes a considerable difficulty in its calculation (especially, for continuous-output channels). Since the bounds in [5, 7] are derived for ensembles of codes, they are probabilistic in their nature; based on concentration arguments, they hold asymptotically in probability 1 as the block length goes to infinity. Based on heuristic statistical mechanics calculations, it was conjectured that the bounds in [7], which hold for general LDPC and LDGM ensembles over MBIOS channels, are tight.

We present in this paper new bounds on the achievable rates and the asymptotic parity-check density of sequences of binary linear block codes. The bounds in [1, 12] and this paper are valid for *every* sequence of binary linear block codes, in contrast to a high probability result which was previously derived for the binary erasure channel (BEC) from density evolution analysis [13]. Shokrollahi proved in [13] that when the codes are communicated over a BEC, the growth rate of the average right degree (i.e., the average degree of the parity-check nodes in a bipartite Tanner graph) is at least logarithmic in terms of the gap to capacity. The statement in [13] is a high probability result, and hence it is not necessarily satisfied for every particular code from this ensemble. Further, it assumes a sub-optimal (iterative) decoding algorithm, where the statements in [1, 12] and this paper are valid even under optimal ML decoding.

In this paper, preliminary material is presented in Section 2, and the theorems are introduced in Section 3. Numerical results are exemplified and explained in Section 4. Finally, we summarize the discussion in Section 5 and present interesting issues which deserve further research. The interested reader is referred to the full paper version [14] which includes the proofs of the theorems in Section 3, discussions and some additional numerical results which were omitted here because of space limitations.

We note that the statements in this paper refer to the case where the parity-check matrices are full rank. Though it seems like a mild requirement for specific linear codes, this poses a problem when considering ensembles of LDPC codes. In the latter case, a parity-check matrix, referring to a randomly chosen bipartite graph with a given pair of degree distributions, may not be full rank. Fortunately, as we later explain in this paper (see Section 4), the statements still hold for ensembles when we replace the code rate with the design rate.

## 2 Preliminaries

We introduce here some definitions and theorems from [1, 12] which serve as a preliminary material for the rest of the paper. Definitions 2.1 and 2.2 are taken from [12, Section 2].

**Definition 2.1 (Capacity-Approaching Codes).** Let $\{\mathcal{C}_m\}$ be a sequence of codes of rate $R_m$, and assume that for every $m$, the codewords of the code $\mathcal{C}_m$ are transmitted with equal probability over a channel whose capacity is $C$. This sequence is said to *achieve a fraction $1 - \varepsilon$ of the channel capacity with vanishing bit error probability* if $\lim_{m \to \infty} R_m = (1-\varepsilon)C$, and if there exists a decoding algorithm under which the average bit error probability of the code $\mathcal{C}_m$ tends to zero in the limit where $m \to \infty$.

**Definition 2.2 (Parity-Check Density).** Let $\mathcal{C}$ be a binary linear code of rate $R$ and block length $n$, which is represented by a parity-check matrix $H$. We define the *density* of $H$, call it $\Delta = \Delta(H)$, as the normalized number of ones in $H$ *per information bit*. The total number of ones in $H$ is therefore equal to $nR\Delta$.

**Definition 2.3 (Log-Likelihood Ratio (LLR)).** Let $p_{Y|X}(\cdot|\cdot)$ be the conditional *pdf* of an arbitrary MBIOS channel. The log-likelihood ratio (LLR) at the output of the

channel is
$$\mathrm{LLR}(y) \triangleq \ln \left( \frac{p_{Y|X}(y|X=0)}{p_{Y|X}(y|X=1)} \right).$$

Throughout the paper, we assume that all the codewords of a binary linear block code are equally likely to be transmitted. Also, we use the notation $h_2(\cdot)$ for the binary entropy function to base 2, i.e., $h_2(\cdot) = -x \log_2(x) - (1-x) \log_2(1-x)$.

**Theorem 2.1 (An Upper Bound on the Achievable Rates for Reliable Communication over MBIOS Channels).** [1, Theorem 2]: Consider a sequence $\{\mathcal{C}_m\}$ of binary linear block codes of rate $R_m$, and assume that their block length tends to infinity as $m \to \infty$. Let $H_m$ be a full-rank parity-check matrix of the code $\mathcal{C}_m$, and assume that $d_{k,m}$ designates the fraction of the parity-check equations involving $k$ variables. Let

$$d_k \triangleq \liminf_{m \to \infty} d_{k,m}, \quad R \triangleq \liminf_{m \to \infty} R_m. \tag{1}$$

Suppose that the transmission of these codes takes place over an MBIOS channel with capacity $C$ bits per channel use, and let

$$w \triangleq \frac{1}{2} \int_{-\infty}^{\infty} \min\bigl(f(y), f(-y)\bigr) \, dy \tag{2}$$

where $f(y) \triangleq p_{Y|X}(y|X=1)$ designates the conditional pdf of the output of the MBIOS channel. Then, a necessary condition for vanishing block error probability as $m \to \infty$ is

$$R \leq 1 - \frac{1-C}{\sum_k \left\{ d_k \ h_2 \left( \frac{1-(1-2w)^k}{2} \right) \right\}}.$$

**Theorem 2.2 (Lower Bounds on the Asymptotic Parity-Check Density with 2-Levels Quantization).** [12, Theorem 2.1]: Let $\{\mathcal{C}_m\}$ be a sequence of binary linear block codes achieving a fraction $1-\varepsilon$ of the capacity of an MBIOS channel with vanishing bit error probability. Denote $\Delta_m$ as the density of a full-rank parity-check matrix of the code $\mathcal{C}_m$. Then, the asymptotic density satisfies

$$\liminf_{m \to \infty} \Delta_m > \frac{K_1 + K_2 \ln \frac{1}{\varepsilon}}{1-\varepsilon} \tag{3}$$

where

$$K_1 = \frac{(1-C) \ln \left( \frac{1}{2 \ln 2} \frac{1-C}{C} \right)}{2C \ \ln \left( \frac{1}{1-2w} \right)}, \quad K_2 = \frac{1-C}{2C \ \ln \left( \frac{1}{1-2w} \right)} \tag{4}$$

and $w$ is defined in (2). For a BEC with erasure probability $p$, the coefficients $K_1$ and $K_2$ in (4) are improved to

$$K_1 = \frac{p \ \ln \left( \frac{p}{1-p} \right)}{(1-p) \ \ln \left( \frac{1}{1-p} \right)}, \quad K_2 = \frac{p}{(1-p) \ \ln \left( \frac{1}{1-p} \right)}. \tag{5}$$

Using standard notation, an ensemble of $(n, \lambda, \rho)$ LDPC codes is characterized by its length $n$, and the polynomials $\lambda(x) = \sum_{i=2}^{\infty} \lambda_i x^{i-1}$ and $\rho(x) = \sum_{i=2}^{\infty} \rho_i x^{i-1}$, where $\lambda_i$ ($\rho_i$) is equal to the probability that a randomly chosen edge is connected to a variable (parity-check) node of degree $i$. The variables (parity-check sets) are represented by the left (right) nodes of a bipartite graph which represents an LDPC code. The design rate of this ensemble is given by

$$R_\mathrm{d} = 1 - \frac{\int_0^1 \rho(x)\,dx}{\int_0^1 \lambda(x)\,dx}.$$

## 3 Main Results

### 3.1 Approach I: Bounds Based on Quantization of the LLR

In this section, we introduce bounds on the achievable rates and the asymptotic parity-check density of sequences of binary linear block codes. The bounds generalize previously reported results in [1] and [12] which were based on a symmetric two-level quantization of the LLR. This is achieved by extending the concept of quantization to an arbitrary integer power of 2; to this end, the analysis relies on the Galois field $\mathrm{GF}(2^d)$. We commence by deriving a lower bound on the conditional entropy of a transmitted codeword given the received sequence.

**Proposition 3.1.** Let $\mathcal{C}$ be a binary linear block code of length $n$ and rate $R$. Let $\mathbf{x} = (x_1, \ldots x_n)$ and $\mathbf{y} = (y_1, \ldots, y_n)$ designate the transmitted codeword and received sequence, respectively, when the communication takes place over an MBIOS channel with conditional pdf $p_{Y|X}(\cdot|\cdot)$. For an arbitrary $d \geq 2$ and $0 \leq l_{2^{d-1}-1} \leq \ldots \leq l_2 \leq l_1 \leq l_0 \triangleq \infty$, let us define the set of probabilities $\{p_s\}_{s=0}^{2^d-1}$ as follows:

$$p_s \triangleq \begin{cases} \Pr\{l_{s+1} < \mathrm{LLR}(Y) \leq l_s \mid X = 0\} & s = 0, \ldots, 2^{d-1} - 2 \\ \Pr\{0 < \mathrm{LLR}(Y) \leq l_{2^{d-1}-1} \mid X = 0\} + \frac{1}{2}\Pr\{\mathrm{LLR}(Y) = 0 \mid X = 0\} & s = 2^{d-1} - 1 \\ \Pr\{-l_{2^{d-1}-1} \leq \mathrm{LLR}(Y) < 0 \mid X = 0\} + \frac{1}{2}\Pr\{\mathrm{LLR}(Y) = 0 \mid X = 0\} & s = 2^{d-1} \\ \Pr\{-l_{2^d-(s+1)} \leq \mathrm{LLR}(Y) < -l_{2^d-s} \mid X = 0\} & s = 2^{d-1} + 1, \ldots, 2^d - 1. \end{cases} \quad (6)$$

For an arbitrary full-rank parity-check matrix of the code $\mathcal{C}$, let $d_k$ denote the fraction of the parity-checks involving $k$ variables. Then, the conditional entropy of the transmitted codeword given the received sequence satisfies

$$\frac{H(\mathbf{X}|\mathbf{Y})}{n} \geq 1 - C - (1-R) \sum_k \left\{ d_k \sum_{\substack{k_0,\ldots,k_{2^{d-1}-1} \\ \sum_i k_i = k}} \binom{k}{k_0, \ldots, k_{2^{d-1}-1}} \cdot \prod_{i=0}^{2^{d-1}-1} (p_i + p_{2^d-1-i})^{k_i} h_2\left(\frac{1}{2}\left[1 - \prod_{i=0}^{2^{d-1}-1} \left(1 - \frac{2p_{2^d-1-i}}{p_i + p_{2^d-1-i}}\right)^{k_i}\right]\right) \right\}.$$

**Theorem 3.1 ("$2^d$-Level Quantization" Lower Bound on the Asymptotic Parity-Check Density of Binary Linear Block Codes).** Let $\{\mathcal{C}_m\}$ be a sequence of binary linear block codes achieving a fraction $1 - \varepsilon$ of the capacity of an MBIOS channel

with vanishing bit error probability. Denote $\Delta_m$ as the density of a full-rank parity-check matrix of the code $\mathcal{C}_m$. Then, the asymptotic density satisfies

$$\liminf_{m \to \infty} \Delta_m > \frac{K_1 + K_2 \ln \frac{1}{\varepsilon}}{1 - \varepsilon}$$

where

$$K_1 = K_2 \ln\left(\frac{1}{2\ln(2)} \frac{1-C}{C}\right), \quad K_2 = -\frac{1-C}{C \ln\left(\sum_{i=0}^{2^{d-1}-1} \frac{(p_i - p_{2^d-1-i})^2}{p_i + p_{2^d-1-i}}\right)}.$$

Here, $d \geq 2$ is an arbitrary integer and the probabilities $\{p_i\}$ are introduced in (6) in terms of $l_1 \geq \ldots \geq l_{2^{d-1}-1} \in \mathbb{R}^+$. The optimal vector of quantization levels $(l_1, \ldots, l_{2^{d-1}-1})$ is given implicitly by the set of $2^{d-1} - 1$ equations

$$\frac{p_{2^d-1-i}^2 + e^{-l_i} p_i^2}{(p_i + p_{2^d-1-i})^2} = \frac{p_{2^d-i}^2 + e^{-l_i} p_{i-1}^2}{(p_{i-1} + p_{2^d-i})^2}, \quad i = 1, \ldots, 2^{d-1} - 1.$$

where such a solution always exists.

**Corollary 3.1 ("$2^d$-Level Quantization" Upper Bound on the Asymptotic Achievable Rates of Sequences of Binary Linear Block Codes).** Let $\{\mathcal{C}_m\}$ be a sequence of binary linear block codes whose codewords are transmitted with equal probability over an MBIOS channel, and suppose that the block length of this sequence of codes tends to infinity as $m \to \infty$. Let $d_{k,m}$ be the fraction of the parity-check nodes of degree $k$ in an arbitrary representation of the code $\mathcal{C}_m$ by a bipartite graph which corresponds to a full-rank parity-check matrix. Then a necessary condition for this sequence to achieve vanishing bit error probability as $m \to \infty$ is that the asymptotic rate $R$ of this sequence satisfies

$$R \leq 1 - \max\left\{(1-C)\left\{\sum_k d_k \sum_{\substack{k_0,\ldots,k_{2^{d-1}-1} \\ \cdot \sum_i k_i = k}} \binom{k}{k_0,\ldots,k_{2^{d-1}-1}} \cdot \prod_{i=0}^{2^{d-1}-1} (p_i + p_{2^d-1-i})^{k_i}\right.\right.$$

$$\left.\left. h_2\left(\frac{1}{2}\left[1 - \prod_{i=0}^{2^{d-1}-1}\left(1 - \frac{2 p_{2^d-1-i}}{p_i + p_{2^d-1-i}}\right)^{k_i}\right]\right)\right\}^{-1}, \frac{2 \sum_{i=2^{d-1}}^{2^d-1} p_i}{1 - \sum_k d_k \left(1 - 2 \sum_{i=2^{d-1}}^{2^d-1} p_i\right)^k}\right\}$$

where $d \geq 2$ is arbitrary, the probabilities $\{p_i\}$ are introduced in (6), and $d_k$ and $R$ are introduced in (1).

## 3.2 Approach II: Bounds without Quantization of the LLR

Similarly to the previous section, we derive bounds on the asymptotic achievable rate and the asymptotic parity-check density of an arbitrary sequence of binary, linear block codes transmitted over an MBIOS channel. As in Section 3.1, the derivation of these two bounds is based on a lower bound on the conditional entropy of a transmitted codeword given the received sequence at the output of an arbitrary MBIOS channel.

**Proposition 3.2.** Let $\mathcal{C}$ be a binary linear code of length $n$ and rate $R$ transmitted over an MBIOS channel. Let $\mathbf{x} = (x_1, x_2, \ldots, x_n)$ and $\mathbf{y} = (y_1, y_2, \ldots, y_n)$ be the transmitted codeword and the received sequence, respectively. For an arbitrary representation of the code $\mathcal{C}$ by a full-rank parity-check matrix, let $d_k$ designate the fraction of the parity-check equations of degree $k$. Then the conditional entropy of the transmitted codeword given the received sequence satisfies

$$\frac{H(\mathbf{X}|\mathbf{Y})}{n} \geq 1 - C - (1-R)\left(1 - \frac{1}{2\ln(2)} \sum_{p=1}^{\infty} \frac{1}{p(2p-1)} \sum_k d_k \left(\int_0^\infty a(l)(1+e^{-l})\tanh^{2p}\left(\frac{l}{2}\right) dl\right)^k\right)$$

where $a(\cdot)$ denotes the conditional pdf of the LLR given that the transmitted symbol is zero.

**Theorem 3.2 ("Un-Quantized" Lower Bound on the Asymptotic Parity-Check Density of Binary Linear Block Codes).** Let $\{\mathcal{C}_m\}$ be a sequence of binary linear codes achieving a fraction $1-\varepsilon$ of the capacity $C$ of an MBIOS channel with vanishing bit error probability. Denote $\Delta_m$ as the density of a full-rank parity-check matrix of the code $\mathcal{C}_m$. Then, the asymptotic density satisfies

$$\liminf_{m \to \infty} \Delta_m \geq \sup_{x \in (0,A]} \frac{K_1(x) + K_2(x) \ln \frac{1}{\varepsilon}}{1-\varepsilon} \qquad (7)$$

where

$$K_1(x) = \frac{1-C}{C} \frac{\ln\left(\frac{\xi(1-C)}{C}\right)}{\ln\left(\frac{1}{x}\right)}, \quad K_2(x) = \frac{1-C}{C} \frac{1}{\ln\left(\frac{1}{x}\right)}.$$

and

$$A \triangleq \int_0^\infty a(l) \frac{(1-e^{-l})^2}{1+e^{-l}} dl, \quad \xi \triangleq \begin{cases} 1 & \text{for a BEC} \\ \frac{1}{2\ln(2)} & \text{otherwise} \end{cases}.$$

**Remark 3.1.** For a BEC with erasure probability $p$, the maximization of the RHS of (7) yields that the maximal value is achieved for $x = 1 - p$ when $\varepsilon < \frac{p}{1-p}$ (otherwise, if $\varepsilon \geq \frac{p}{1-p}$, the lower bound is useless as it becomes non-positive). Hence, the lower bound on the asymptotic parity-check density stated in Theorem 3.2 coincides with the bound for the BEC in [12, Eq. (3)]. This lower bound was demonstrated in [12, Theorem 2.3] to be tight. This is proved by showing that the sequence of right-regular LDPC ensembles of Shokrollahi [13] is optimal in the sense that it achieves (up to a small additive coefficient) the lower bound on the asymptotic parity-check density for the BEC.

**Remark 3.2.** The lower bound on the parity-check density in Theorem 3.2 is uniformly tighter than the one in [12, Theorem 2.1] (except for the BSC and BEC where they coincide). For a proof of this claim, the reader is referred to [14, Appendix D.2].

**Corollary 3.2 ("Un-Quantized" Upper Bound on the Asymptotic Achievable Rates of Sequences of Binary Linear Block Codes).** Let $\{\mathcal{C}_m\}$ be a sequence of binary linear block codes whose codewords are transmitted with equal probability over an MBIOS channel, and assume that the block lengths of these codes tend to infinity as $m \to \infty$. Let $d_{k,m}$ be the fraction of the parity-check nodes of degree $k$ for arbitrary representations of the codes $\mathcal{C}_m$ by bipartite graphs which corresponds to a full-rank

parity-check matrix. Then a necessary condition on the achievable rate ($R$) for obtaining vanishing bit error probability as $m \to \infty$ is

$$R \leq 1 - \frac{1-C}{1 - \frac{1}{2\ln(2)} \sum_{p=1}^{\infty} \left\{ \frac{1}{p(2p-1)} \sum_k d_k \left( \int_0^{\infty} a(l)(1+e^{-l}) \tanh^{2p}\left(\frac{l}{2}\right) dl \right)^k \right\}}$$

where $d_k$ and $R$ are introduced in (1).

## 4 Numerical Results

We present here numerical results for the information-theoretic bounds in Section 3. As expected, they significantly improve the numerical results presented in [1, Section 4] and [12, Section 4]. This improvement is attributed to the fact that, in contrast to [1, 12], in the derivation of the bounds in this paper, we do not perform a two-level quantization of the LLR which in essence converts the arbitrary MBIOS channel (whose output may be continuous) to a BSC. Throughout this section, we assume transmission of the codes over the binary-input AWGN channel.

We note that the statements in Sections 2 and 3 refer to the case where the parity-check matrices are full rank. Though it seems like a natural requirement for specific linear codes, this poses a problem when considering ensembles of LDPC codes. In the latter case, a parity-check matrix, referring to a randomly chosen bipartite graph with a given pair of degree distributions, may not be full rank. Considering ensembles of LDPC codes, it follows from the proofs in [14, Sections 3 and 4] that the statements stay valid with the following modifications: the actual code rate $R$ of a code which is randomly picked from the ensemble is replaced by the design rate ($R_{\mathrm{d}}$) of the ensemble, and $\{d_k\}$ becomes the degree distribution of the parity-check nodes referring to the original bipartite graph which represents a parity-check matrix, possibly not of full rank. By doing this modification, the bound becomes looser when the asymptotic rate of the codes is strictly above the design rate of the ensemble. In light of this modification, we note that the fraction $d_k$ of nodes of degree $k$ is calculated in terms of the degree distribution $\rho(\cdot)$ by the equation

$$d_k = \frac{\rho_k}{k} \frac{1}{\int_0^1 \rho(x)\, dx} \,.$$

This conclusion about the possible replacement of the code rate with the design rate in case that the parity-check matrices are not full rank was also noted in [1, p. 2439].

### 4.1 Thresholds of LDPC Ensembles under ML Decoding

Table 1 provides bounds on the thresholds of Gallager's regular LDPC ensembles under ML decoding. It also gives an indication on the inherent loss in performance due to the sub-optimality of MPI decoding.

The bounds on the achievable rates in [1] and Corollaries 3.1 and 3.2 provide lower bounds on the $\frac{E_\mathrm{b}}{N_0}$ thresholds under ML decoding. For Gallager's regular LDPC ensembles, the gap between the thresholds under ML decoding and the exact thresholds under the sum-product decoding algorithm (calculated using density-evolution analysis) are rather large. For this reason, we also compare the lower bounds on the $\frac{E_b}{N_0}$ thresholds under ML decoding with upper bounds on the $\frac{E_b}{N_0}$ thresholds which rely on "typical pairs decoding" [3]; an upper bound on the $\frac{E_b}{N_0}$ thresholds under an arbitrary sub-optimal decoding algorithm (e.g., "typical pairs decoding") also forms an upper bound on these thresholds under optimal ML decoding. It is shown in Table 1 that the gap between the thresholds under iterative decoding and the bounds for ML decoding (see the columns referring to the DE threshold and the upper bound based on "typical pairs decoding") is rather large. This is attributed to the sub-optimality of belief propagation decoding for regular LDPC ensembles. On the other hand, it is also demonstrated in Table 1 that the gap between the upper and lower bounds on the thresholds under ML decoding is much smaller. For example, according to the numerical results in Table 1, the inherent loss in the asymptotic performance due to the sub-optimality of belief propagation for Gallager's ensemble of $(4,6)$ regular LDPC codes (whose design rate is $\frac{1}{3}$ bits per channel use) ranges between 2.097 and 2.137 dB.

| LDPC | Capacity Limit | 2-Level Bound [1] | 4-Level Bound | 8-Level Bound | Un-Quantized Lower Bound | Upper Bound [3] | DE Threshold |
|---|---|---|---|---|---|---|---|
| (3,6) | +0.187 dB | +0.249 dB | +0.332 dB | +0.361 dB | +0.371 dB | +0.673 dB | 1.110 dB |
| (4,6) | −0.495 dB | −0.488 dB | −0.472 dB | −0.463 dB | −0.463 dB | −0.423 dB | 1.674 dB |
| (3,4) | −0.794 dB | −0.761 dB | −0.713 dB | −0.694 dB | −0.687 dB | −0.510 dB | 1.003 dB |

Table 1: Comparison of thresholds for Gallager's ensembles of regular LDPC codes transmitted over the binary-input AWGN channel. The 2-level lower bound on the threshold of $\frac{E_b}{N_o}$ refers to ML decoding, and is based on [1, Theorem 1] (see also [12, Table II]). The 4-level, 8-level and un-quantized lower bounds apply to ML decoding, and are based on Corollaries 3.1 and 3.2. The upper bound on the threshold of $\frac{E_b}{N_o}$ holds under 'typical pairs' decoding [3] (and hence, also under ML decoding), and the density evolution (DE) thresholds are based on density evolution for MPI decoding [9].

Similarly to Table 1, numerical results referring to ensembles of irregular LDPC codes which closely approach the capacity limit over the binary-input AWGN channel are shown in [14, Section 5].

The plots in Figure 1 compare different lower bounds on the $\frac{E_b}{N_0}$-threshold under ML decoding of right-regular LDPC ensembles. The plots refer to a right degree of 6 (left plot) or 10 (right plot). The following lower bounds are depicted in these plots: the Shannon capacity limit, the 2-level quantization lower bound in [1, Theorem 1], the 4 and 8-level quantization bounds of the LLR in Section 3.1, and finally, the bound in Section 3.2 where no quantization of the LLR is performed. It can be observed from the two plots in Figure 1 that the range of code rates where there exists a visible improvement with the new lower bounds depends on the degree of the parity-check nodes. In principle, the larger the value of the right-degree is, then the improvement obtained by these bounds is more pronounced starting from a higher rate code rate (e.g., for a right degree of 6 or 10, the improvement obtained by the new bounds is observed for code rates starting from 0.35 and 0.55 bits per channel use, respectively).

## 4.2 Lower Bounds on the Asymptotic Parity-Check Density

The lower bound on the parity-check density derived in Theorem 3.2 enables to assess the tradeoff between asymptotic performance and asymptotic decoding complexity (per iteration) of an MPI decoder. This bound tightens the lower bound on the asymptotic parity-check density derived in [12, Theorem 2.1]. Fig. 2 compares these bounds for codes of rate $\frac{1}{2}$ (left plot) and $\frac{3}{4}$ (right plot) where the bounds are plotted as a function of $\frac{E_b}{N_0}$. It can be observed from Fig. 2 that as $\frac{E_b}{N_0}$ increases, the advantage of the bound in Theorem 3.2 over the bound in [12, Theorem 2.1] diminishes. This follows from the fact that as the value of $\frac{E_b}{N_0}$ is increased, the two-level quantization of the LLR used in [1] and [12, Theorem 2.1] better captures the true behavior of the MBIOS channel. It is also reflected in this figure that as $\varepsilon$ tends to zero (i.e., when the gap to capacity vanishes), the slope of the bounds becomes very sharp. This is due to the logarithmic behavior of the bounds.

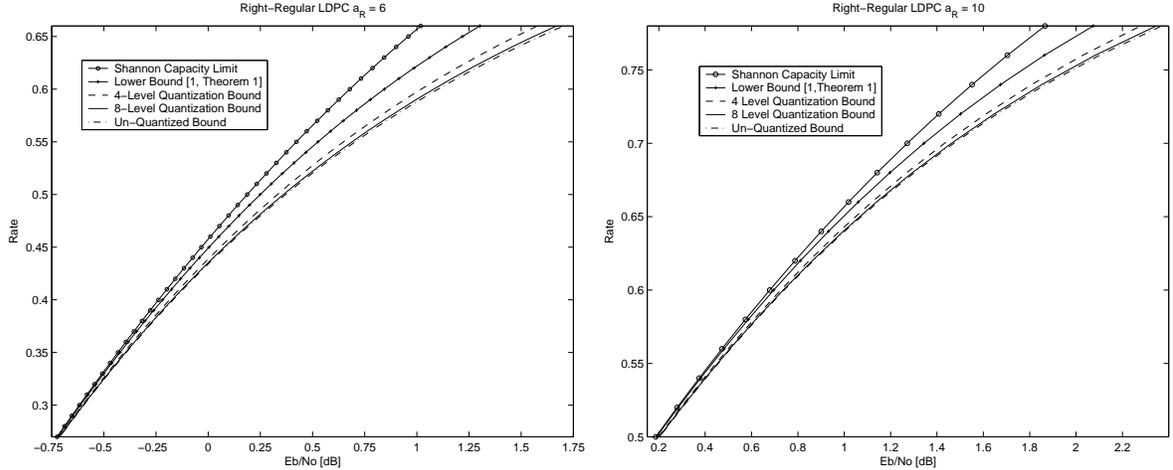

Figure 1: Comparison between different lower bounds on the threshold under ML decoding for right-regular LDPC ensembles with $a_R = 6$ (left plot) and $a_R = 10$ (right plot). The transmission takes place over the binary-input AWGN channel.

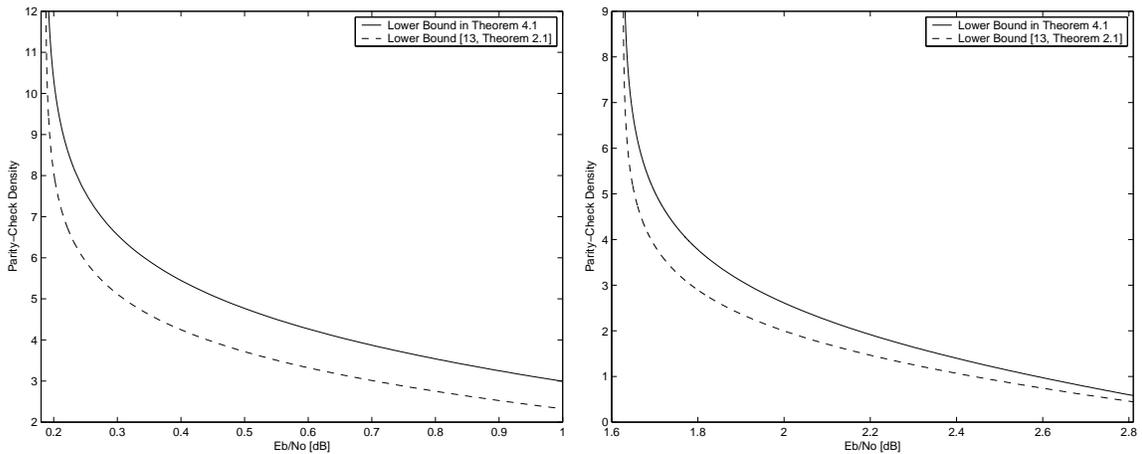

Figure 2: Comparison between lower bounds on the asymptotic parity-check density of binary linear block codes transmitted over a binary-input AWGN channel. The dashed line refers to [12, Theorem 2.1], and the solid line refers to Theorem 3.2. The left and right plots refer to code rates of $\frac{1}{2}$ and $\frac{3}{4}$, respectively. The Shannon capacity limit for these code rates corresponds to $\frac{E_b}{N_0}$ of 0.187 dB and 1.626 dB, respectively.

# 5  Summary and Outlook

We present improved lower bounds on the asymptotic density of parity-check matrices and upper bounds on the achievable rates of binary linear block codes transmitted over memoryless binary-input output-symmetric (MBIOS) channels. The improvements are w.r.t. the bounds given in [1, 12]. The information-theoretic bounds are valid for *every* sequence of binary linear block codes, in contrast to high probability results which follow from probabilistic analytical tools (e.g., density evolution (DE) analysis under iterative decoding). The bounds hold under optimal ML decoding, and hence, they hold in particular under any sub-optimal decoding algorithm. We apply the bounds to ensembles of low-density parity-check (LDPC) codes. The significance of the bounds is the following: firstly, by comparing the new upper bounds on the achievable rates with thresholds provided by DE analysis, we obtain rigorous bounds on the inherent loss in performance of various LDPC ensembles. This degradation in the asymptotic performance is due to the sub-optimality of MPI decoding (as compared to optimal ML decoding). Secondly, the parity-check density can be interpreted as the complexity per iteration under MPI decoding. Therefore, by tightening the reported lower bound on the asymptotic parity-check density [12, Theorem 2.1], the new bounds provide better insight on the tradeoff between the asymptotic performance and the asymptotic decoding complexity of iteratively decoded LDPC codes. Thirdly, a new lower bound on the bit error probability of binary linear block codes presented in [14, Corollary 4.2] tightens the reported lower bound in [12, Theorem 2.5].

The comparison between the quantized and un-quantized bounds gives insight on the effect of the number of quantization levels of the LLR (even if they are chosen optimally) on the achievable rates, as compared to the ideal case where no quantization is done. The results of such a comparison are shown in Table 1, and indicate that the improvement in the tightness of the bounds when more than 8 levels of quantization are used (in case the quantization levels are optimally determined) is marginal.

The bounds on the thresholds of LDPC ensembles under optimal ML decoding depend only on the degree distribution of their parity-check nodes and their design rate. For a given parity-check degree distribution ($\rho$) and design rate ($R$), they provide an indication on the inherent gap to capacity which is independent of the choice of $\lambda$ (as long as the pair of degree distributions ($\lambda, \rho$) yield the design rate $R$). Therefore, the bounds are not expected to be tight for LDPC ensembles with a given pair of degree distributions ($\lambda, \rho$). The numerical results shown in Table 1 indicate, however, that these bounds are useful for assessing the inherent gap to capacity of LDPC ensembles. As a topic for further research, it is suggested to examine the possibility of tightening the bounds for specific ensembles by explicitly taking into account the exact characterization of $\lambda$.

The lower bound on the asymptotic parity-check density in [12, Theorem 2.1] and its improvements in Section 3 grow like the log of the inverse of the gap (in rate) to capacity. The result in [12, Theorem 2.3] shows that a logarithmic growth rate of the parity-check density is achievable for Gallager's regular LDPC ensemble under ML decoding when transmission takes place over an arbitrary MBIOS channel. These results show that for any iterative decoder which is based on the representation of the codes by Tanner graphs, there exists a tradeoff between asymptotic performance and complexity which cannot be surpassed. Recently, it was shown in [8] that better tradeoffs can be achieved by allowing more complicated graphical models; for the particular case of the binary erasure channel (BEC), the encoding and the decoding complexity of properly designed codes on graphs remain bounded as the gap to capacity vanishes. To this end, Pfister, Sason and Urbanke consider in [8, Theorems 1 and 2] ensembles of irregular repeat-accumulate codes which involve punctured bits, and allow in this way a sufficient number of state nodes in the Tanner graph representing the codes. This surprising result is considered in [8, Theorem 4], by a derivation of an information-theoretic lower bound on the decoding complexity of randomly punctured codes on graphs whose transmission takes place over MBIOS channels. The approach for the derivation of the bounds in [8, Theorem 4] relies on the analysis in [12, Theorem 2.1]. It is suggested to tighten the lower bounds in [8, Theorem 4] by relying on the approach used to prove Theorem 3.2 in [14] (as compared to the derivation of the bound in [8, Theorem 4] which relies on the analysis in [12, Theorem 2.1]). The interested reader is referred to [14], including proofs of the theorems, discussions and more numerical results that were omitted here for the sake of brevity.

## Acknowledgment

The second author wishes to acknowledge Rüdiger Urbanke for stimulating discussions during the preparation of the work in [12] which motivated the research in this paper. The authors acknowledge Andrea Montanari for his correspondence regarding [7].